\documentclass[a4paper]{jpconf}
\usepackage{graphicx}
\usepackage{multirow}

\begin{document}

\title{Quenching Factor consistency across several NaI(Tl) crystals}

\author{D.~Cintas$^1$, P.~An$^{2,3}$, C.~Awe$^{2,3}$, P.~S.~Barbeau$^{2,3}$, E.~Barbosa de Souza$^4$, S.~Hedges$^{2,3}$, J.~H.~Jo$^4$, M.~Mart\'inez$^{1,5}$, R.~H.~Maruyama$^4$, L.~Li$^{2,3}$, G.~C.~Rich$^3$, J.~Runge$^{2,3}$, M.~L.~Sarsa$^1$, W.~G.~Thompson$^4$}

\address{$^1$ Centro de Astropart\'{\i}culas y F\'{\i}sica de Altas Energ\'{\i}as (CAPA), Universidad de Zaragoza, 50009 Zaragoza, Spain}
\address{$^2$ Department of Physics, Duke University, Durham, NC 27708, USA}
\address{$^3$ Triangle Universities Nuclear Laboratory, Durham, NC 27708, USA}
\address{$^4$ Department of Physics and Wright Laboratory, Yale University, New Haven, CT 06520, USA}
\address{$^5$ Fundaci\'on ARAID, Av. de Ranillas 1D, 50018 Zaragoza, Spain}

\begin{abstract}
Testing the DAMA/LIBRA annual modulation result independently of dark matter particle and halo models has been a challenge for twenty years. Using the same target material, NaI(Tl), is required and presently two experiments, ANAIS-112 and COSINE-100, are running for such a goal. A precise knowledge of the detector response to nuclear recoils is mandatory because this is the most likely channel to find the dark matter signal. The light produced by nuclear recoils is quenched with respect to that produced by electrons by a factor that has to be measured experimentally. However, current quenching factor measurements in NaI(Tl) crystals disagree within the energy region of interest for dark matter searches. To disentangle whether this discrepancy is due to intrinsic differences in the light response among different NaI(Tl) crystals, or has its origin in unaccounted for systematic effects will be key in the comparison among the different experiments. We present measurements of the quenching factors for five small NaI(Tl) crystals performed in the same experimental setup to control systematics. Quenching factor results are compatible between crystals and no clear dependence with energy is observed from 10 to 80~keVnr.
\end{abstract}

\section{Introduction}
For about twenty years the DAMA/LIBRA collaboration has been reporting the observation of an annual modulation in the detection rate which is compatible with dark matter \cite{Rita}. This observation is in strong tension with results from experiments using different targets, but a model independent test was lacking until recently. Such a goal requires using the same target material, NaI(Tl), and it is being pursued by two experiments in data-taking phase: ANAIS-112 [2-3] and COSINE-100 [4-5].

A good knowledge of the detector response to nuclear recoils is crucial for DAMA/LIBRA signal testing because many DM candidates are expected to scatter off nuclei. Nuclear recoils (NR) are less efficient than electrons in producing scintillation, for the same energy deposition. Such quenching of the scintillation signal for NR translates into a different energy scale. The ratio of the number of scintillation photons produced by NR and electrons is called quenching factor (QF), and is required to convert electron equivalent energies into nuclear recoil energies. While electron equivalent energy scale is usually well known, because most of the experiments are calibrated with gamma sources, uncertainties in QF for many detectors/nuclei are large. Figure 1 shows some of the results for the QF of sodium nuclei in NaI(Tl) crystals for different measurements [6-11], showing discrepancies with the DAMA/LIBRA value.

The measurement of the QF of sodium in NaI(Tl) crystals presented here was done in the Advanced Neutron Calibration Facility at Triangle Universities Nuclear Laboratory (TUNL), at Duke University (North Carolina, US). Five small NaI(Tl) crystals grown from different powders and in different ingots have been measured in the same setup in order to study possible QF dependencies on crystal properties and possible systematic effects. A timing-based trigger was implemented in order to avoid threshold effects. 

\section{Experimental setup}

The Advanced Neutron Calibration Facility at TUNL provides a quasi-mono energetic neutron beam, which crosses a collimator made of polyethylene and induces nuclear recoils in the NaI(Tl) crystals. A lead wall shields the detectors from beam-induced radiation. The energy distribution of the neutrons has been measured using a time-of-flight detector. The nuclear recoil energies produced in the NaI(Tl) detector have been determined by the angle of the scattered neutron, identified by the interaction in one of several liquid scintillator detectors, called Backing Detectors (BDs), as seen in Figure 2.

\begin{figure}[h]
\begin{center}
\begin{minipage}{19pc}
\includegraphics[width=19pc]{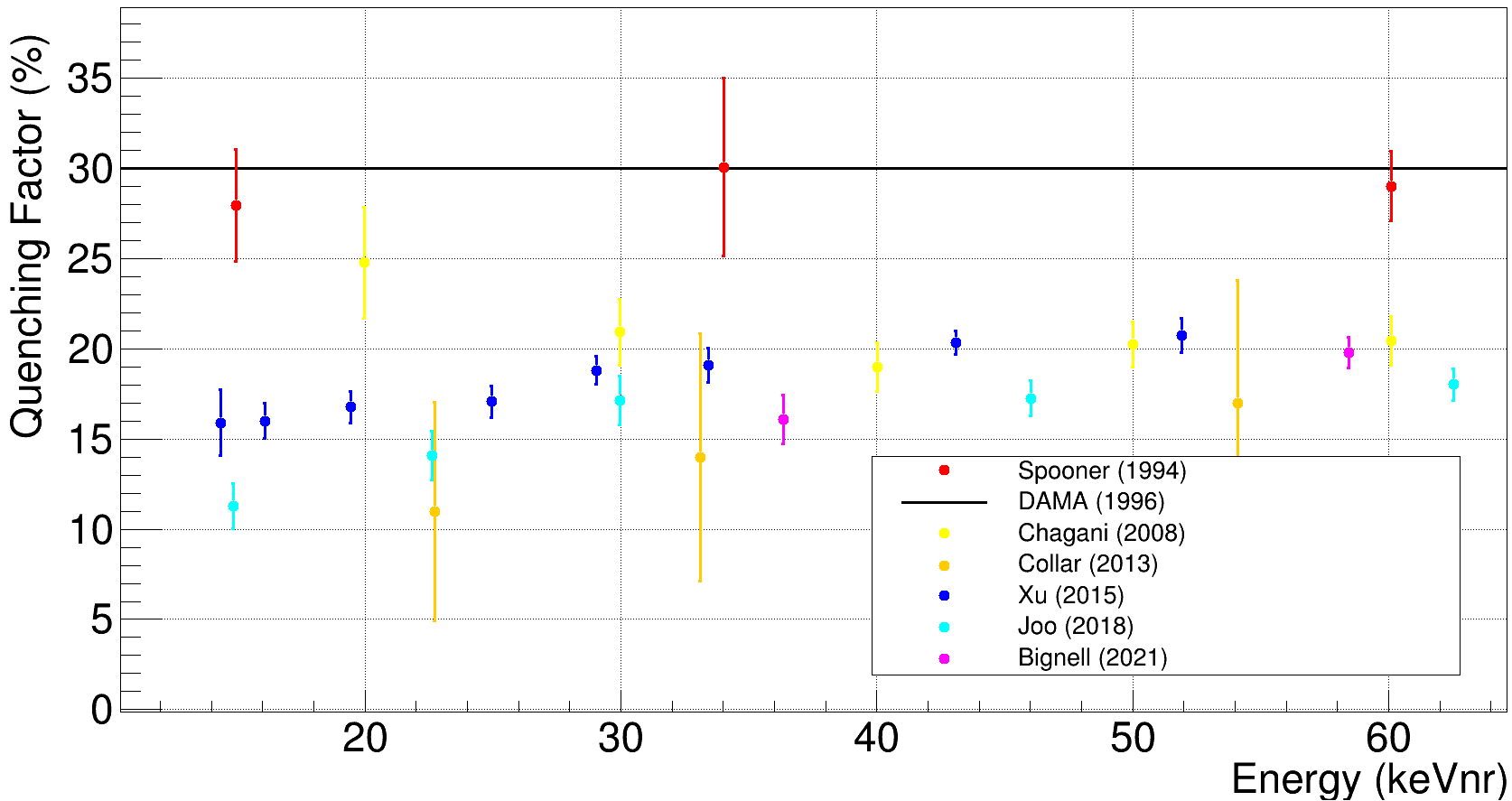}
\caption{\label{QF}Quenching factor measurements for sodium nuclei in NaI(Tl) crystals [6-11].}
\end{minipage}\hspace{2pc}%
\begin{minipage}{16pc}\vspace{1.1pc}
\includegraphics[width=16pc]{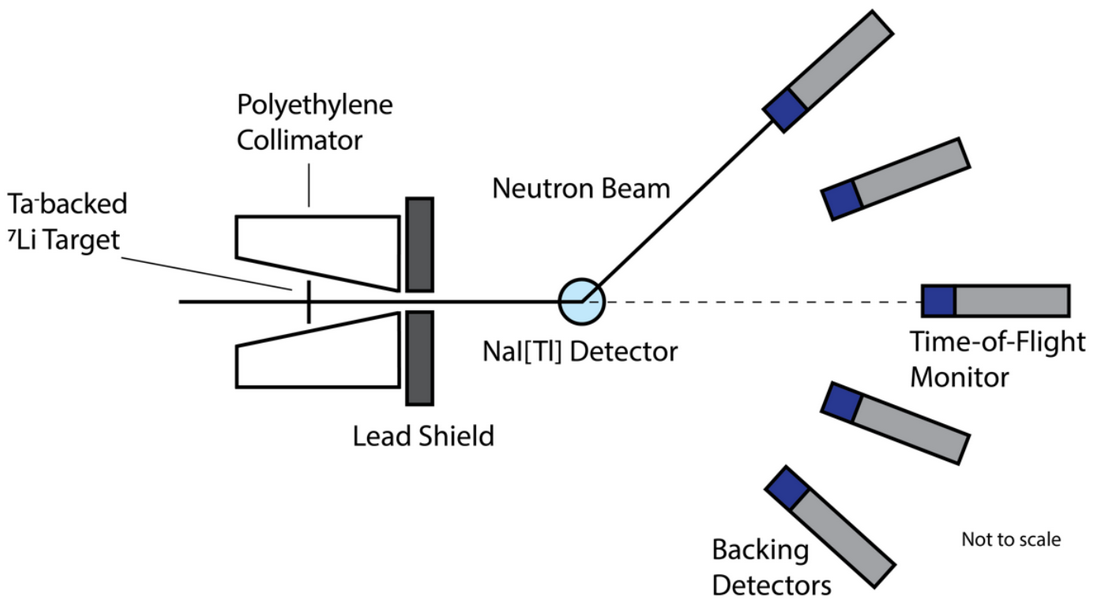}
\centering\caption{\label{tunl}Schematic plot of the experimental setup.}
\end{minipage}
\end{center}
\end{figure}

To reduce systematics, the same PMT has been used for the measurement of all the crystals, which are conveniently small to minimize multiple scattering. Table \ref{table} summarizes the properties of the five crystals measured. These measurements were done in two different runs: on August and October 2018, with neutron mean energies of 958 and 982 keV, respectively.

\begin{table}[h]
\begin{center}
\begin{tabular}[h]{|c|c|c|c|c|c|p{4cm}|}
	\hline 
	Number & Powder & Diameter & Group & Run & Neutron energy\\ 
	\hline 
	1 & WIMP-Scint-III & \multirow{3}{*}{1"} & \multirow{3}{*}{Yale} & \multirow{2}{*}{August '18} & \multirow{2}{*}{\parbox{8em}{$\mu=958\pm5keV$\\ $\sigma = 4 \pm 3 keV$}} \\ 
	\cline{1-2}
	2 & WIMP-Scint-II & & & &  \\ 
	\cline{1-2}\cline{5-6}
	3 & WIMP-Scint-I & & & \multirow{3}{*}{October '18} & \multirow{3}{*}{\parbox{8em}{$\mu=982\pm7keV$\\ $\sigma = 7 \pm 5 keV$}} \\ 
	\cline{1-4}
	4 & AS Standard & \multirow{2}{*}{0.5"} & \multirow{2}{*}{Zaragoza} & &\\ 
	\cline{1-2}
	5 & WIMP-Scint-III & & & &\\ 
	\hline
\end{tabular}
\caption{\label{table}Summary of crystal and run properties. $\mu$ is the mean and $\sigma$ the standard deviation of the neutron energy distributions.}
\end{center}
\end{table}

Each crystal was measured on-beam for 20-36 hours. The acquisition of the waveforms was triggered by an event in one of the 18 BDs. This allows to define the signal region in the NaI(Tl) crystal (2800ns integration window) by a time-offset from the global trigger defined by the BDs. Every eight hours crystals were rotated in order to reduce possible channeling effects. For each crystal, three off-beam measurements were carried out: a measurement of the background spectrum and $^{133}$Ba calibration of the NaI(Tl) crystal, and $^{137}$Cs calibration of the BDs.

The rate in the BDs is dominated by beam-induced gamma events. However, neutrons can be identified using the time of flight to the triggered BD (calculated using a beam monitoring) and a pulse shape parameter (defined as the fraction of the pulse area corresponding to the pulse tail in the BD waveform).

\section{Geant4 simulation}

A Geant4 Monte Carlo simulation of the full experimental setup has been developed considering all the relevant elements of the setup:

\begin{itemize}
    \item The crystal and its complete housing
    \item The 18 BDs
    \item The lead shield
    \item The time of flight detector (present only in the specific neutron beam energy measurements)
    \item The neutron source (defined as point-like, emitting neutrons within a 2.4º angle and with the energy distribution evaluated from the time-of-flight measurement).
\end{itemize}

Two simulations have been performed:  under neutron irradiation and under exposure to a $^{133}$Ba source. The former allows to obtain the probability distribution function (PDF) for the recoil energies in the NaI(Tl) crystal in coincidence with any of the BDs, which will be used to derive the QF, as it will be described in section 5.

\section{NaI(Tl) calibration}

\begin{figure}[h]
\begin{center}
\begin{minipage}{18pc}
\includegraphics[width=18pc]{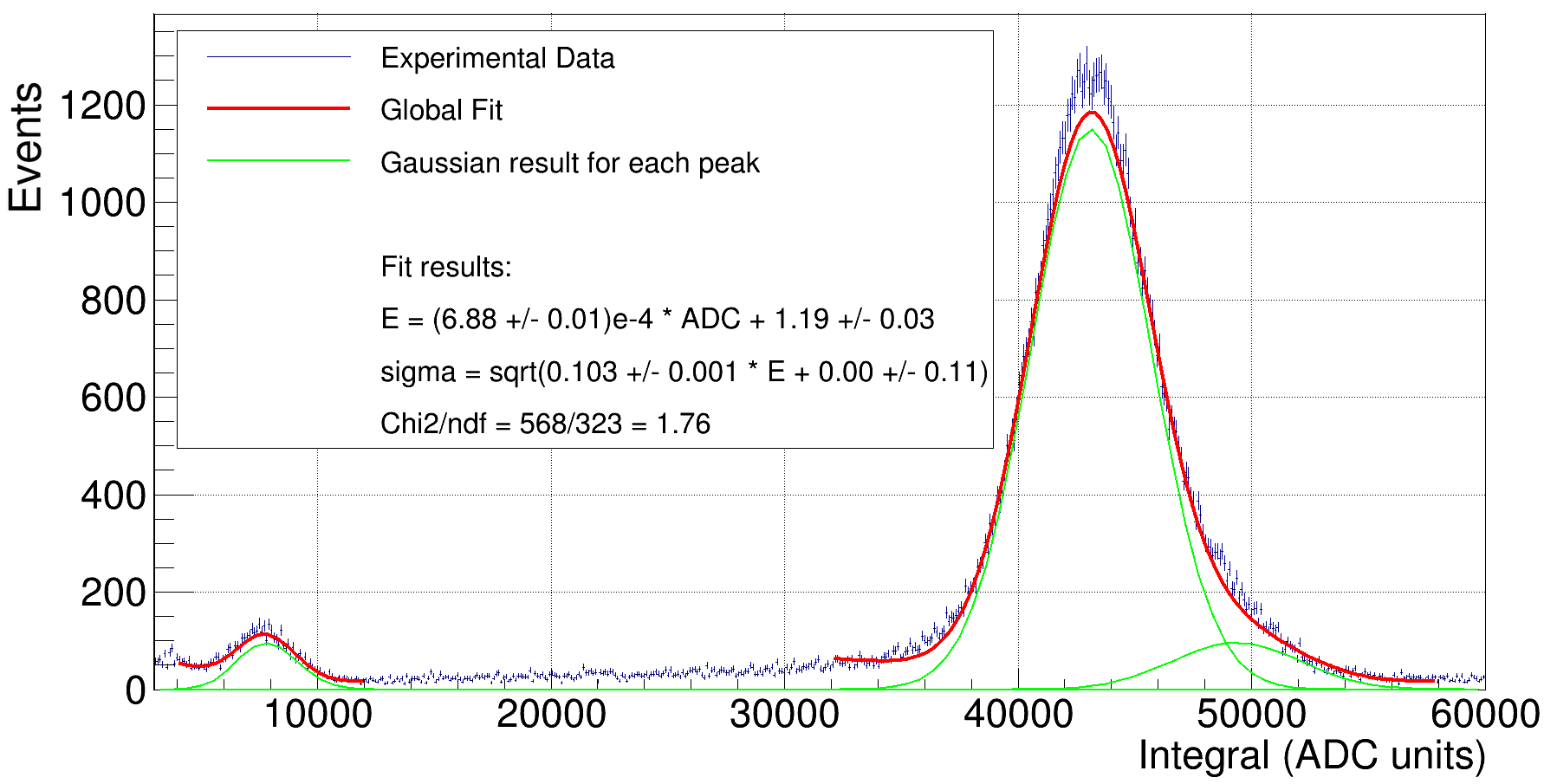}
\centering\caption{\label{calFit}$^{133}$Ba spectrum measured by crystal 1 and global fit (see text).}
\end{minipage}\hspace{2pc}%
\begin{minipage}{17pc}\vspace{0.75pc}
\includegraphics[width=17pc]{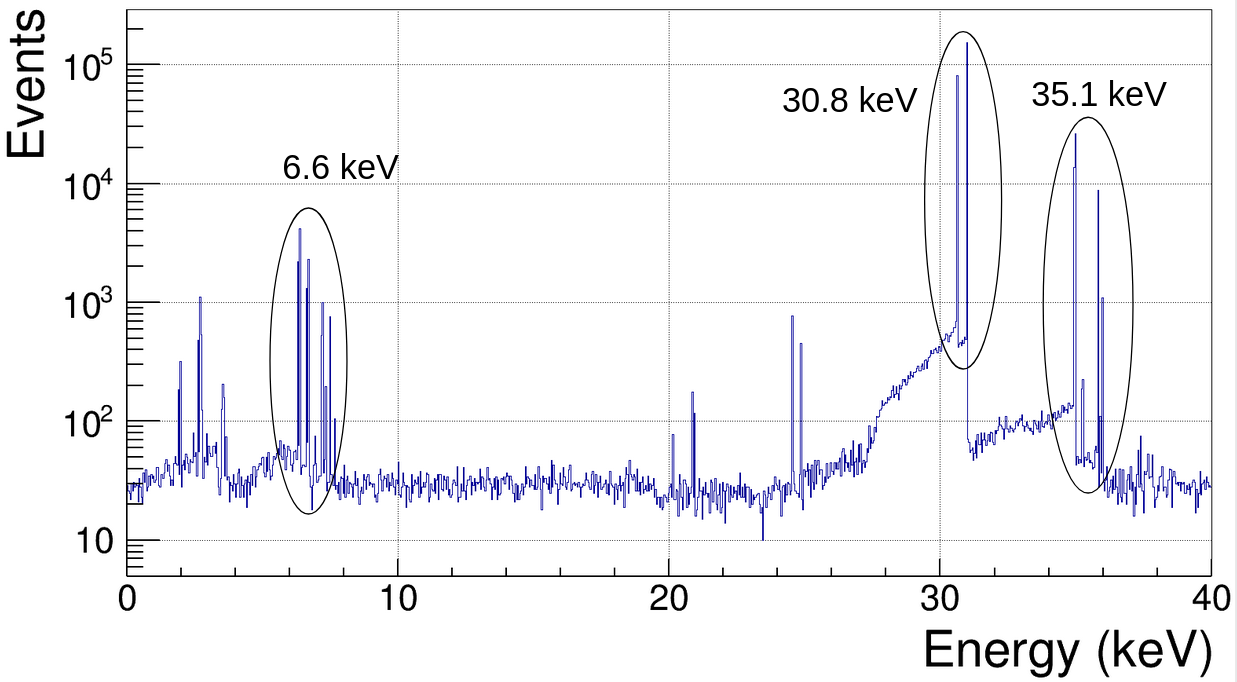}
\centering\caption{\label{calSim}Simulation of the energy deposited in the NaI(Tl) crystal exposed to the $^{133}$Ba source.}
\end{minipage}
\end{center}
\end{figure}

Figure~\ref{calFit} shows the experimental spectrum for the $^{133}$Ba calibration of crystal 1. In order to identify the energies of the peaks visible in the spectrum, a MC simulation has been used. Figure~\ref{calSim} shows the result of the simulation of the energy deposited in the NaI(Tl) crystal exposed to the $^{133}$Ba source. Because of the limited energy resolution of the NaI(Tl), we have assigned to the observed peaks in figure~\ref{calFit} the weighted mean of the dominant lines observed in the simulated spectrum.

The experimental spectrum (Fig.~\ref{calFit}) has been fitted to three gaussian functions having as free parameters the coefficients of a linear relation between ADC and energy and those of the function relating energy resolution and energy, $\sigma(E) = \sqrt{a\cdot E + b}$. The resulting electron equivalent energy calibration is robust between 5 and 40 keVee, where we observe neutron-induced NR.

\section{Quenching Factor calculation procedure and results}

\begin{figure}[h]
\begin{center}
\begin{minipage}{37pc}
\includegraphics[width=37pc]{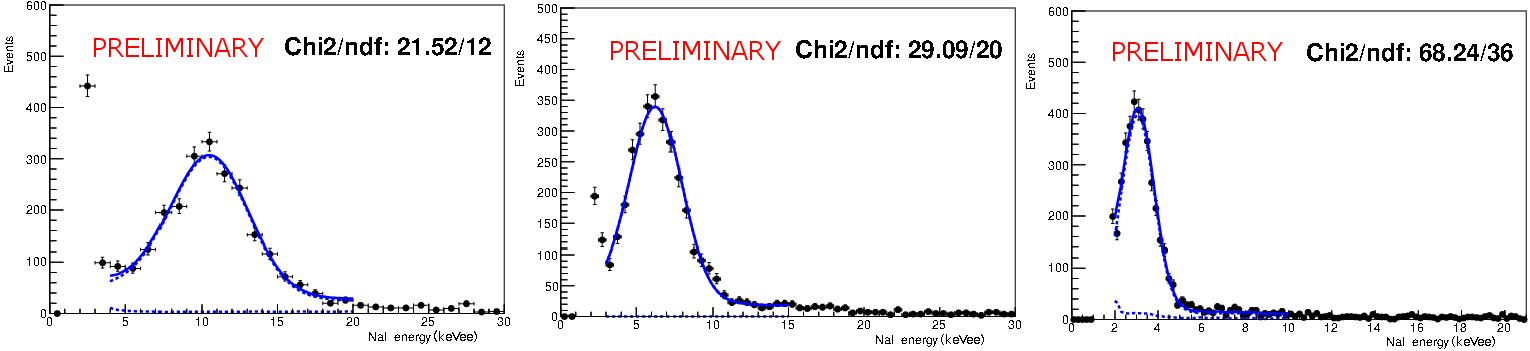}
\caption{\label{qf_fit}Examples of the NR measured distributions for crystal 1 for three scattering angles (from left to right: 69, 52 and 33 degrees), and the fit to the PDF derived from the simulation.}
\end{minipage}
\end{center}
\end{figure}

For the events that have fired a particular BD and therefore correspond to a certain NR energy in the NaI(Tl), we convert the signal into electron equivalent energy using the calibration described in section 4. This spectrum is fitted to the PDF calculated from the MC simulation, which is convolved with a resolution function and converted into electron equivalent energy through a scaling factor, which is the sought QF. Both the resolution function parameters and the QF are free parameters in the fit. Figure~\ref{qf_fit} shows three examples of those fits for crystal 1.

\begin{figure}[h]
\begin{center}
\begin{minipage}{17pc}
\includegraphics[width=17pc]{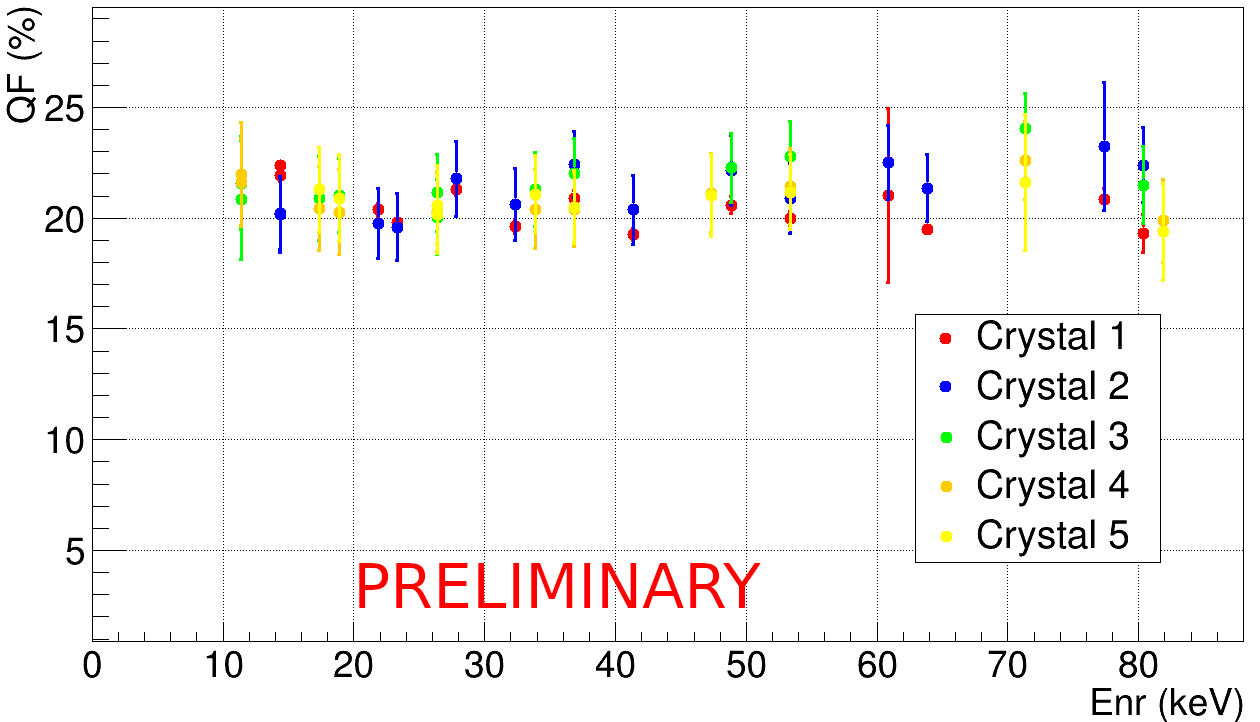}
\centering\caption{\label{results_Ba133}QF results using $^{133}$Ba calibration (section 5).}
\end{minipage}\hspace{2pc}%
\begin{minipage}{17pc}
\includegraphics[width=17pc]{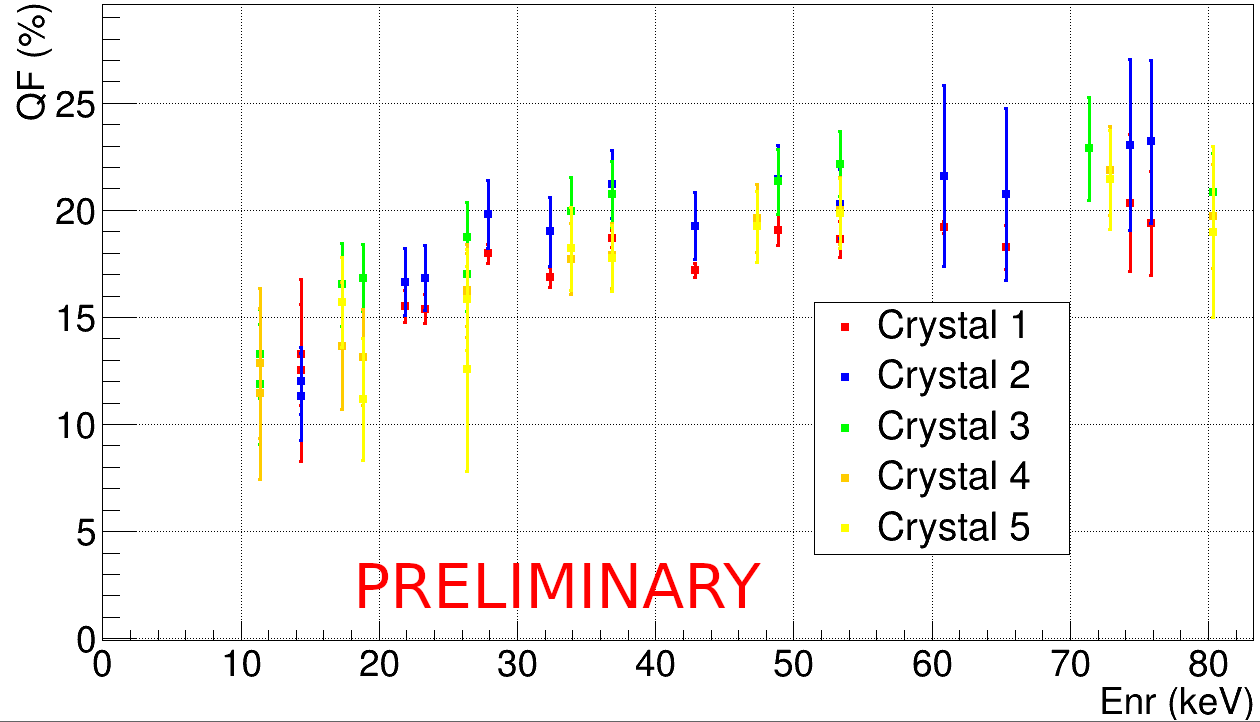}
\centering\caption{\label{results_57}QF results using the 57.6 keV line and assuming proportionality.}
\end{minipage}
\end{center}
\end{figure}

Figure~\ref{results_Ba133} summarizes the results of the QFs for all the crystals. The error bars include the effects of the systematic uncertainties in the BD, NaI(Tl) and neutron source positions and also from the electron equivalent energy calibration. The Na QF results of all crystals are compatible within their uncertainties and do not depend on energy. The mean QF for the five crystals is $20.9 \pm 0.3 \%$. This result is still preliminary.

The QF results are very sensitive to the electron equivalent energy calibration. When using instead of the $^{133}$Ba calibration explained before, a proportional response of the NaI(Tl) taking as reference the 57.6 keV line (emitted following neutron-$^{127}I$ inelastic scattering), we obtain the QF results shown in Figure~\ref{results_57}. The apparent decrease in QF at low energies (as reported by other experiments, e.g. [8-11]) disappears if we use the $^{133}$Ba calibration in all of the five measured crystals, leading us to conclude that non-proportional behaviour in the detector response has to be taken into consideration.

Pure Iodine recoils are below our energy threshold, therefore we use the inelastic scattering gamma line at 57.6 keV and look for a shift of this line with respect to the nominal position due to the addition of the quenched iodine NR energy. Only upper limits could be obtained with this method. Crystals 1, 4 and 5 had poor statistics or poor resolution, depending on the crystal, so they have not been considered. For crystals 2 and 3 the upper limits at 90\% C.L. are 9.4\% and 8.2\% for energies of 11 and 13 keVnr, respectively.

\section{Conclusions}

The QF of five NaI(Tl) crystals has been measured at TUNL. Compatible values within uncertainties have been obtained for all the crystals. No clear dependence with energy is observed from 10 to 80 keVnr, with a mean value of $20.9 \pm 0.3 \%$. It has been observed that recent disagreements in the QF measurements could be attributed to the non-proportional response of the detectors. This analysis is still ongoing and further studies are required to fully understand the QF in NaI(Tl). This preliminary result, confirming the consistency of the QF values across different NaI(Tl) crystals, supports direct model-independent testing of the DAMA/LIBRA annual modulation result by ANAIS-112 and COSINE-100 experiments.

\section*{Acknowledgments}
This work has been financially supported by the Spanish Ministerio de Econom\'ia y Competitividad and the European Regional Development Fund (MINECO-FEDER) under Grant No. FPA2017-83133-P; the Ministerio de Ciencia e Innovaci\'on – Agencia Estatal de Investigaci\'on under Grant No. PID2019-104374GB-I00; the Consolider-Ingenio 2010 Programme under Grants No. MultiDark CSD2009-00064 and No. CPAN CSD2007-00042 and the LSC Consortium. This work is supported in part by the U.S. Department of Energy under grant DE-FG02-97ER41033 and by NFS Grants No. PHY-1913742, and DGE1122492.


\section*{References}
\begin{thebibliography}{9}
\bibitem{Rita} R. Bernabei et al. 2020 Prog. Part. Nucl. Phys 114, 103810
\bibitem{ANAIS-1} J. Amar\'e et al. 2019 Phys. Rev. Lett. 123, 031301
\bibitem{ANAIS-2} J. Amar\'e et al. 2021 Phys. Rev. D 103 102005
\bibitem{COSINE-1} G. Adhikari et al. 2018 Nature 564, 83
\bibitem{COSINE-2} G. Adhikari et al. 2019 Phys. Rev. Lett. 123 031302
\bibitem{Spooner} N. Spooner et al. 1994 Phys. Lett. B 321, 156
\bibitem{Chagani} H.  Chagani et  al. 2008  JINST 3 (06), 06003
\bibitem{Collar} J. I. Collar et al. 2013 Phys. Rev. C88, 035806
\bibitem{Xu} J. Xu, et  al. 2015 Phys. Rev. C92, 015807
\bibitem{Joo}  H.  Joo et al. 2019 Astropart. Phys. 108, 50
\bibitem{Bignell}  L. J. Bignell et al. 2021 JINST 16 (07), 07034


\end{thebibliography}
\end{document}